%% file: paper.tex
\documentclass{LMCS}

\input{macros}

\usepackage{epsfig,amsmath,xspace,bcprules,enumerate,boxedminipage}
\usepackage{hyperref}

\def\doi{3 (2:1) 2007}
\lmcsheading%
{\doi}
{1--20}
{}
{}
{May\phantom{.}~30, 2006}
{Apr.~18, 2007}
{}   

\begin{document}

\title{Predicate Abstraction via Symbolic Decision Procedures\rsuper *}





\author[S.~K.~Lahiri]{Shuvendu K. Lahiri\rsuper a}	
\address{{\lsuper a}Microsoft Research, Redmond, WA 98052}	
\email{\{shuvendu,tball\}@microsoft.com}  

\author[T.~Ball]{Thomas Ball\rsuper a}
\address{\vskip -6 pt}	

\author[B.~Cook]{Byron Cook\rsuper b}
\address{{\lsuper b}Microsoft Research, Cambridge, United Kingdom}	
\email{bycook@microsoft.com}  



\titlecomment{{\lsuper *}An earlier version of the paper appeared in the proceedings of the Computer Aided Verification conference, 2005}

\input{abstract}

\maketitle

\date{\today}

\input{intro}
\input{setup}

\input{fastf}
\input{impl}
\input{concl}

\bibliography{refs}
\bibliographystyle{alpha}


\end{document}

%% file: macros.tex
\def\eff{\mathcal{F}}
\def\gee{\mathcal{G}}

\def\DP{\it DP}
\def\SDP{\it SDP}

\def\false{{\tt false}}
\def\true{{\tt true}}

\def\c2bp{\textsc{C2bp}\xspace}

\def\mynot{\neg}
\def\myimplies{\Rightarrow}

\newcommand{\comment}[1]{}

\newcommand{\bnf}{\mathrel{::=}}
\newcommand{\vbar}{\mathbin{|}}
\newcommand{\lvbar}{|\;}

\newcommand{\va}[1]{\rm {\it #1}}
\newcommand{\mva}[1]{\mbox{\va{#1}}}
\newcommand{\syntaxform}[1]{\mva{#1}}
\newcommand{\sterm}{\syntaxform{term}}
\newcommand{\sformula}{\syntaxform{formula}}
\newcommand{\satomformula}{\syntaxform{atomic-formula}}
\newcommand{\svar}{\syntaxform{variable}}
\newcommand{\sfunction}{\syntaxform{function-symbol}}
\newcommand{\spredicate}{\syntaxform{predicate-symbol}}

\newcommand{\tempexpr}[1]{t[{#1}]}

\newcommand{\tempexprset}[1]{{\it tset}\_{\{#1\}}}
\newcommand{\derivgoal}{\mathrel{:-}}
\newcommand{\derivDepth}[2]{{\it derivDepth}_{#1}(#2)}
\newcommand{\maxDerivDepth}[2]{{\it maxDerivDepth}_{#1}(#2)}

\newcommand{\evalExpr}[2]{\it eval(#1,#2)}

\newcommand{\varset}[1]{\it V_{\it #1}}
\newcommand{\clause}[1]{\it cl_{#1}}

\newcommand{\neccVars}[1]{\it NeccVars(#1)}

\newcommand{\eClass}[1]{\it EC(#1)}

%% file: abstract.tex
\begin{abstract}
We present a new approach for performing predicate abstraction based
on symbolic decision procedures. Intuitively, a symbolic decision
procedure for a theory takes a set of predicates in the theory and
symbolically executes a decision procedure on all the subsets over the
set of predicates.  The result of the symbolic decision procedure is
a shared expression (represented by a directed acyclic graph) that
implicitly represents the answer to a predicate abstraction query.

We present symbolic decision procedures for the logic of Equality and
Uninterpreted Functions (EUF) and Difference logic (DIFF) and show
that these procedures run in pseudo-polynomial (rather than
exponential) time.  We then provide a method to construct symbolic
decision procedures for simple mixed theories (including the two
theories mentioned above) using an extension of the Nelson-Oppen
combination method. We present preliminary evaluation of our Procedure
on predicate abstraction benchmarks from device driver verification in
SLAM.

%
%
\end{abstract}

%% file: intro.tex
\section{Introduction}
\label{seC:intro}

Predicate abstraction is a technique for automatically 
creating  finite abstract models of finite and infinite state
systems~\cite{graf-cav97}. The method has been widely used in
abstracting finite-state models of programs in 
SLAM~\cite{ball-pldi01} and numerous
other software verification projects~\cite{henzinger-popl02,magic-tse04}. 
It has also been used for synthesizing loop 
invariants~\cite{flanagan-popl02} and verifying distributed
protocols~\cite{das-cav99,lahiri-cav03a}. 

The fundamental operation in predicate abstraction can be summarized
as follows: Given a set of predicates $P$ describing some set of
properties of the system state, and a formula $e$, compute the weakest
Boolean formula $\eff_P(e)$ over the predicates $P$ that implies
$e$\footnote{The dual of this problem, which is to compute the
  strongest Boolean formula $\gee_P(e)$ that is implied by $e$, can be
  expressed as $\neg \eff_P(\neg e)$.}. Most implementations of
predicate abstraction~\cite{graf-cav97,ball-pldi01}
construct $\eff_P(e)$ by collecting the set of cubes (a conjunction of
the predicates or their negations) over $P$ that imply $e$. The
implication is checked using a first-order theorem prover. This method
may require making a very large ($2^{|P|}$ in the worst case) number
of calls to a theorem prover and can be expensive.

We propose a new way to perform predicate abstraction based on {\it
  symbolic decision procedures}. A symbolic decision procedure for a
theory $T$ ($\SDP_T$) takes sets of predicates $G$ and $E$ and
symbolically executes a decision procedure for $T$ on $G' \cup \{\neg
e \;|\; e \in E\}$\footnote{Throughout this paper, we interpret a set
  of expressions to be a conjunction over the expressions in the set.}, for all the subsets $G'$ of $G$.
The output of $\SDP_T(G,E)$ is a shared expression (an expression
where common subexpressions can be shared) representing those subsets 
$G' \subseteq G$, for which $G' \cup \{\neg e \;|\; e \in E\}$ is
unsatisfiable. We show that such a procedure can be used to 
compute $\eff_P(e)$ for performing predicate abstraction.

We present symbolic decision procedures for the logic of Equality and
Uninterpreted Functions(EUF) and Difference logic (DIF) and show that
these procedures run in polynomial and pseudo-polynomial time
respectively, and therefore produce compact shared expressions. We
provide a method to construct $\SDP$ for a combination of two simple
theories $T_1 \cup T_2$ (including EUF + DIF), by using an extension
of the Nelson-Oppen combination~\cite{nelson-jacm80} method.  We use
Binary Decision Diagrams (BDDs)~\cite{bryant-tc86} to construct
$\eff_P(e)$ from the shared representations efficiently in practice.

We present a preliminary evaluation of our procedure on predicate
abstraction benchmarks from device driver verification in SLAM, and
show that our method outperforms existing methods for doing predicate
abstraction.

The rest of the paper is organized as follows:
Section~\ref{sec:related} describes related work in predicate
abstraction techniques. Section~\ref{sec:setup} describes the
background concepts including predicate
abstraction. Section~\ref{sec:fastf} describes symbolic decision
procedures, and instantiates it for two different theories (EUF and
DIF). Section~\ref{sec:combine-sdp} describes a framework for
modularly combining the SDPs for two theories that satisfy certain
requirements, using an extension of the Nelson-Oppen combination
method. Section~\ref{sec:impl} describes the implementation and the
experimental evaluation of our technique. Finally, we present the
conclusions and future work in Section~\ref{sec:concl}. 

\input{related}




\comment{
Predicate abstraction is a technique for automatically creating
abstract models of finite and infinite state
systems~\cite{graf-cav97}. Let $E = \{ e_1, \cdots, e_n \}$ be a set
of predicates over the state space of a system. Given another
predicate $e$, the basic problem in predicate abstraction is to find
the weakest boolean function $F_E(e)$ over the predicates $E$ that
implies $e$.  The function $F_E(e)$ may be represented as a
disjunction $c_1 \vee \cdots \vee c_m$, where each $c_j$ (a so-called
``minterm'') is a conjunction of literals $l_{j_1} \wedge \cdots
\wedge l_{j_n}$ such that each $l_{j_k} \in \{ e_k, \neg e_k \}$ and
such that each $c_j$ implies $q$.  The naive approach to compute
$F_E(e)$ is to enumerate each of the $2^n$ minterms and query a
theorem prover to see if the formula $c_j \implies e$ is valid, for
each minterm $c_j$. If so, the minterm $c_j$ is included in $F_E(e)$.
Otherwise, it is not.

The problem of computing $F_E(e)$ appears to be inherently exponential
in $n$ (the number of predicates in $E$).  Various techniques have
been considered for optimizing the predicate abstraction
process~\cite{XYZ}, but sacrifice precision to gain efficiency.
Further, most implementations suffer from the fact that a very large
number of independent theorem prover calls have to be made to compute
$\eff_E(e)$. {\bf [TODO: this needs more work]}

We present a new approach for performing predicate abstraction based
on {\it symbolic decision procedures}.  To understand our approach, it
is instructive to consider what a decision procedure for a decidable
logic $L$ does. Given a set of predicates $E$ in the logic $L$ and a
goal predicate $g$, a decision procedure $DP_L$ checks whether or not
$(\bigwedge_{e \in E} e) \wedge \neg g$ is satisfiable. The logic $L$
can be characterized by a set of axioms and inference rules, which
$DP_L$ applies to resolve whether or not the input formula is
satisfiable.

A symbolic decision procedure takes a set of predicates $E$ (in logic
$L$) and a goal predicate $g$ and symbolically simulates the action of
$DP_L$ to check the satisfiability for every subset $E'$ of $E$ with
respect to $g$. In the process, it constructs an expression to denote
all the possible computations of $DP_L$. With $E$ predicates, we can
have $2^n$ such subsets, and our procedure simulates the behavior of
$DP_L$ on each of them. We show that our process works for theories
for which a decision procedure can be constructed using a Nelson-Oppen
combination of other decision procedures. The proof and all the
arguments depend on this observation and is not dependent on whether
the computations are stored explicitly or as a directed acyclic graph
(DAG). For simple theories (equality with uninterpreted functions and
difference constraints), our process builds a compact
(pseudo-polynomial sized) data structure that can be effectively
evaluated to produce $F_E(e)$ in time proportional to the size of
$F_E(e)$.
            
This process is analogous to the symbolic simulation of a transition
system with a transition relation and an initial condition.  In our
case, the inference rules of the logic serve as the transition
relation and the set of predicates $E$ and axioms of the logic serve
as the initial state condition. We use the max-derivation depth (which
exists for the given decidable logics) to prove convergence, which is
similar to symbolic simulating until the diameter of the state space
of a system.



Experimental evaluation of the technique for SLAM. We also extend 
our method to handle non-atomic goals by first converting the goal
to an equivalent CNF expression. 

\comment{
The contributions of the paper:
\begin{itemize}
\item Efficiently represent the derivations for atomic predicates in 
a theory. 
\item For mixed theories with certain restriction, we can use Nelson-Oppen
framework to perform the task modularly. Convergence criteria 
dependent upon particular representation. 
\item Provide two refinements of this technique with complementary
strengths on simple theories. 
\begin{enumerate}
\item With {\bf dag} representation, we can provide a pseudo polynomial
sized shared representation of the derivations. Can also provide
efficient enumeration using BDDs in practice.  
\item With the {\bf worklist} algorithm, we can provide an output
sensitive (?) technique. The technique is {\it incremental}. Suited for
Nelson-Oppen combination and Boolean program construction. 
\end{enumerate}
\item Experimental evaluation
\begin{enumerate}
\item Results on predicate abstraction queries from SLAM. Compare with
Simplify-based method and SAT-based quantification. 
Data with both approaches (size of shared expr, time) etc.
\item Effect on SLAM loop (?). 
\end{enumerate}
\item Extend it to handle queries with Boolean structure. 

\end{itemize}
}
}

%% file: related.tex
\subsection{Related Work}
\label{sec:related}


Several techniques have been suggested to improve the performance of
predicate abstraction. The techniques can be broadly classified into
three categories: In the first category, we classify methods that
treat the decision procedures as a ``black box'', and attempt to
minimize the number of decision procedure calls during predicate
abstraction. The second category consists of methods that use a
quantifier elimination procedure to perform predicate
abstraction. Finally, there are techniques that do not compute the
most precise abstract directly; instead, they rely on counterexamples
or proofs in the overall verification process to refine the
abstraction. In the following paragraphs, we describe these techniques
in more details.

The techniques that aim to reduce the number of calls to the theorem
prover or decision procedure are mostly based on enumerating cubes
over $P$ in an increasing order of their size.  Das et
al.~\cite{das-cav99} enumerates cubes over a tree, after fixing the
order of predicates that appear in any path to the leaves. If a cube
is found unsatisfiable, then all its sub-cubes (represented by the
subtree) are pruned off. This method may require $2^{|P|+1}$ calls to
the theorem prover in the worst case. Saidi and
Shankar~\cite{saidi-cav99} relaxes the order on the predicates, and
enumerate all possible cubes ($3^{|P|}$ of them) over the
predicates. Flanagan and Qadeer~\cite{flanagan-popl02} provide an
algorithm that searches over the $2^{|P}|$ clauses (disjunction of
cubes over the predicates or their negations) of size $|P|$, but
attempts to greedily grow the clause (by dropping literals) when such
a clause is implied by the formula $e$. Their technique requires
$|P|.2^{|P|}$ theorem prover calls in the worst case. Other techniques
sacrifice precision to gain efficiency, by only considering cubes of
some fixed length~\cite{ball-pldi01}. All these techniques may require
an exponential number of theorem prover calls in the worst case, and
demonstrate worst case behavior in practice. However, more
importantly, since these queries are not incremental, the state of the
prover has to be reset across each call, precluding any learning
across calls.



Alternately, predicate abstraction can be formulated as a quantifier
elimination problem.  Lahiri et al.~\cite{lahiri-cav03a} and Clarke et
al.~\cite{clarke-fmsd04} perform predicate abstraction by reducing the
problem of computing $\eff_P(e)$ to Boolean quantifier
elimination. The former method first transforms a first-order
quantifier elimination problem into Boolean quantifier elimination by
encoding first-order formulas into Boolean formulas; the latter
assumes all variables are propositional. The method
in~\cite{lahiri-cav03a} first converts the quantifier-free first-order
formula to a Boolean formula such that the translation preserves the
set of satisfying assignments of the Boolean variables in the original
formula. Both these techniques use incremental Boolean Satisfiability
(SAT) techniques~\cite{clarke-fmsd04,mcmillan-cav02} to perform the
Boolean quantifier elimination.  These techniques have the benefit
that the large number of calls to the theorem prover is avoided, and
learning can be used to prune away the search space in the SAT
solver. However, the translation from a first-order formula to a
Boolean formula can result in a loss of structure (since the
arithmetic operations are encoded as bitwise operations), and make the
translation inefficient.  Namjoshi and Kurshan~\cite{namjoshi-cav00}
also proposed using quantifier elimination for first-order logic
directly to perform predicate abstraction --- however many theories
(such as the theory of Equality with Uninterpreted Functions) do not
admit quantifier elimination.


Most of the above approaches use decision procedures or SAT
solvers as ``black boxes'', at best in an incremental fashion,
to perform predicate abstraction. We believe that having a 
customized procedure for predicate abstraction can help improve
the efficiency of predicate abstraction on large problems.

Finally, there are a set of techniques to avoid computing the most
precise abstraction upfront, and refine it only based on failed proof
attempts in the verification tool. Das and Dill~\cite{das-lics01} and
subsequently Ball et al.~\cite{ball-tacas04} use counterexamples to
refine the predicate abstraction incrementally. Jhala and
McMillan~\cite{jhala-cav05} use {\it interpolants} to refine the
predicate abstraction. It is not clear if it is always preferable to
compute the abstraction incrementally. But, we have observed that the
refinement loop can often becomes the main bottleneck in these
techniques (for example in SLAM), and limits the scalability of the
overall system~\cite{ball-tacas04}.

%% file: setup.tex
\section{Setup}
\label{sec:setup}

Figure~\ref{fig:syntax}  defines the syntax of a 
quantifier-free fragment of first-order logic.
An expression in the logic can either be a $\sterm$ or a $\sformula$. 
A $\sterm$ can either be a variable or an application of a function
symbol to a list of terms. 
A $\sformula$ can be the constants $\true$ or $\false$ or an atomic
formula or Boolean combination of other formulas. Atomic formulas
can be formed by an equality between terms or by an application of a 
predicate symbol to a list of terms.

\begin{figure}[t]
\begin{eqnarray*}
\sterm & \bnf & \svar \vbar \sfunction(\sterm,\ldots,\sterm) \\
\satomformula & \bnf & \sterm = \sterm \vbar 
\spredicate(\sterm,\ldots,\sterm) \\
\sformula & \bnf & \true \vbar \false \vbar \satomformula\\
& \lvbar & \sformula \wedge \sformula \vbar \sformula \vee \sformula
\vbar  \neg \sformula 
\end{eqnarray*}
\caption{\label{fig:syntax} Syntax of a quantifier-free fragment of first-order logic.}
\end{figure}

The function and predicate symbols can either be {\it uninterpreted}
or can be defined by a particular theory. For instance, the theory
of integer linear arithmetic defines the function-symbol ``+'' to be 
the addition function over integers and ``$<$'' to be the comparison
predicate over integers. If an expression involves function or 
predicate symbols from multiple theories, then it is said to be an
expression over {\it mixed} theories. 

A formula $F$ is said to be {\it satisfiable} if it is possible to assign
values to the various symbols in the formula from the domains associated
with the theories to make the formula \true. A formula is {\it valid}
if $\neg F$ is not satisfiable (or unsatisfiable). We say a formula 
$A$ {\it implies}  a formula $B$ ($A \myimplies B$) if and only if 
$(\neg A) \vee B$ is valid. 

We define a {\it shared expression} to be a Directed Acyclic Graph (DAG)
representation of an expression 
where common subexpressions can be shared, by using
names to refer to common subexpressions. For example,
the intermediate variable $t$ refers to the expression $e_1$ in
the shared expression ``${\bf let}~~t=e_1~~{\bf in}~~(e_2 \wedge t) \vee (e_3 \wedge \neg t)$''.

\subsection{Predicate Abstraction}
\label{sec:pred-abs}

A {\it predicate} is an atomic formula or its 
negation\footnote{We always use the term ``predicate symbol''
(and not ``predicate'') to refer to symbols like ``$<$''.}. 
If $G$ is a set of predicates, then
we define $\widetilde{G} \doteq \{\neg g \;|\; g \in G \}$,
to be the set containing the negations of the predicates in $G$.
We use the term ``predicate'' in a general sense to refer to any
atomic formula or its negation and should not be confused to 
only mean the set of predicates that are used in predicate abstraction.

\begin{defi}
For a set of predicates $P$, a {\it literal} $l_i$ over $P$ is either a 
predicate $p_i$ or $\neg p_i$, where $p_i \in P$. 
A {\it cube} $c$ over $P$ is a conjunction of  literals.
A {\it clause} $\clause{}$ over $P$ is a disjunction of  literals.
Finally, a {\it minterm} over $P$ is a cube with $|P|$ literals, and 
exactly one of $p_i$ or $\neg p_i$ is present in the cube. 
\end{defi}

Given a set of predicates $P \doteq \{p_1,\ldots,p_n\}$
and a formula $e$, the main operation in
predicate abstraction involves constructing the {\it weakest} Boolean
formula $\eff_P(e)$ over $P$ such that $\eff_P(e) \myimplies e$. 
%
%
%
%
The expression $\eff_P(e)$ can be  expressed as the set 
of all the minterms over $P$ that imply $e$:
\begin{equation}
\eff_P(e) = \bigvee \{c \;|\; c \textrm{ is a minterm over } P \textrm{ and } c \textrm{ implies } e\}
\label{eqn:def_f}
\end{equation}

\begin{prop} For a set of predicates $P$ and a formula $e$, the following statements are true:
\begin{enumerate}
\item $\eff_P(\mynot e) \myimplies \mynot \eff_P(e)$,
\item  $\eff_P(e_1 \wedge e_2) \Leftrightarrow \eff_P(e_1) \wedge \eff_P(e_2)$, and
\item $\eff_P(e_1) \vee \eff_P(e_2) \myimplies \eff_P(e_1 \vee e_2)$
\end{enumerate}
\label{prop:F_properties}
\end{prop}

\begin{proof}
These properties follow very easily from the definition of 
$\eff_P$. 

We know that $\eff_P(e) \myimplies e$, by the definition of 
$\eff_P(e)$. By contrapositive rule, $\mynot e \myimplies \mynot \eff_P(e)$. But
$\eff_P(\mynot e) \myimplies \mynot e$. Therefore, 
$\eff_P(\mynot e) \myimplies \mynot \eff_P(e)$. 

To prove the second equation, we  prove that 
(i) $\eff_P(e_1 \wedge e_2) \myimplies (\eff_P(e_1) \wedge \eff_P(e_2))$, and
(ii) $(\eff_P(e_1) \wedge \eff_P(e_2)) \myimplies \eff_P(e_1 \wedge e_2)$. 
Since $e_1 \wedge e_2 \myimplies e_i$ (for $i \in \{1,2\}$), 
$\eff_P(e_1 \wedge e_2) \myimplies \eff_P(e_i)$. Therefore
$\eff_P(e_1 \wedge e_2) \myimplies (\eff_P(e_1) \wedge \eff_P(e_2))$.
On the other hand, $\eff_P(e_1) \myimplies e_1$ and
$\eff_P(e_2) \myimplies e_2$, $\eff_P(e_1) \wedge \eff_P(e_2) \myimplies
e_1 \wedge e_2$. Since $\eff_P(e_1 \wedge e_2)$ is the weakest expression 
that implies $e_1 \wedge e_2$, $\eff_P(e_1) \wedge \eff_P(e_2) \myimplies
\eff_P(e_1 \wedge e_2)$.

To prove the third equation, 
note that $\eff_P(e_1) \vee \eff_P(e_2) \myimplies e_1 \vee e_2$
and $\eff_P(e_1 \vee e_2)$ is the weakest expression that implies 
$e_1 \vee e_2$.  

\end{proof}

The operation $\eff_P(e)$ does not distribute over disjunctions.
Consider the example where
$P \doteq \{x \not= 5 \}$ and  $e \doteq  x < 5 \vee x > 5$.
In this case, $\eff_P(e) = x \not= 5$. However
$\eff_P(x<5) = \false$ and $\eff_P(x>5) = \false$ and thus
$\left(\eff_P(x<5) \vee \eff_P(x>5)\right)$ is not the same as
$\eff_P(e)$.

The above properties suggest that one can adopt a two-tier approach
to compute $\eff_P(e)$ for any formula $e$:
\begin{enumerate}
\item Convert $e$ into an {\it equivalent} Conjunctive Normal Form
  (CNF), which comprises of a conjunction of clauses, i.e., $e \equiv
  (\bigwedge_i \clause{i})$.  
\item For each clause $\clause{i} \doteq 
(e_1^i \vee e_2^i \ldots \vee e_m^i)$, 
compute $r_i \doteq \eff_P(\clause{i})$ and return
$\eff_P(e) \doteq \bigwedge_i r_i$. 
\end{enumerate}




To obtain an equivalent CNF form, one cannot introduce auxiliary
variables (to keep the size of the resulting formula linear in the
size of the input formula), as is typically done during an
equisatisfiable CNF translation. These auxiliary variables introduced
have to be existentially quantified out to obtain an equivalent
formula.  In our case, the CNF representation of the formula can be
exponentially large compared to the original formula.  However, we can
use recent techniques to obtain the CNF form lazily, by a method
proposed by McMillan~\cite{mcmillan-cav02}.

For the rest of hte paepr, we focus here on computing
$\eff_P(\bigvee_{e_i \in E} e_i)$ when $e_i$ is a predicate. Unless
specified otherwise, we always use $e$ to denote $(\bigvee_{e_i \in E}
e_i)$, a disjunction of predicates in the set $E$ in the sequel.




\comment{
\subsection{Shared Representation of $\eff_P(e)$}
\begin{figure}[t]
\begin{center}
\epsfig{figure=../figs/egraph_example.eps,width=2.5in}
\end{center}
\caption{\label{fig:egraph}An e-graph representing the set of equalities
$\{ (a=b), (a=x), (x=b), (b=c), (b=y), (y=c), (c=d), (c=z), (z=d) \}$.}
\end{figure}

Consider an example where
$P \doteq \{ (a=b), (a=x), (x=b), (b=c), (b=y), (y=c), 
(c=d), (c=z), (z=d) \}$  and $e \doteq (a = d)$.
The graph shown in Figure~\ref{fig:egraph} represents the set of
predicates in $P$, where an edge between vertices $u$ and $v$
corresponds to the predicate $u = v$ in $P$.

It is not hard to see that each simple path between $a$ and $d$ in the 
graph represents a cube in $\eff_P(e)$.  For example, the path 
$a-b-c-d$ represents the cube $(a=b) \wedge (b=c) \wedge (c=d)$, which
implies $a=d$.  Moreover, $\eff_P(e)$ contains precisely those cubes that
correspond to the simple paths from $a$ to $d$ in this graph.  
In this case, we desire an efficient way to encode 
the set of all simple paths (eight, in all) between $a$ and $d$
to compute $\eff_P(e)$.  

Note that the number of such paths can be exponential in the size of the 
graph ($2^{n}$ paths for series-parallel graphs as above with $n$ triangles). 
However, one can compute a shared representation of $\eff_P(a=d)$ 
in polynomial time (and thus representable in polynomial space). One such 
representation of $\eff_P(a=d)$ can be:
\begin{verbatim}
      let t_{a=b} := a=b | (a=x & x=b) in
      let t_{a=c} := (t_{a=b} & b=c) | (t_{a=b} & b=y & y=c) in
      let t_{a=d} := (t_{a=c} & c=d) | (t_{a=c} & c=z & z=d) in
      t_{a=d}
\end{verbatim}
where $\tempexpr{}$ denotes a temporary variable to hold intermediate
expressions\footnote{For this specific case, one can represent the expression for
$\eff_P(a=d)$ as 
{\tt (a=b | (a=c \& c=b)) \& (b=c | (b=y \& b=c)) \& (c=d | (c=z \& z=d))}
without any temporary variables, but this is not true for more 
complicated graphs.}.

We will refer to an expression that uses names to refer to intermediate
expressions as a {\it shared expression}. Implicitly, we can imagine 
a Directed Acyclic Graph (DAG) representation of the expression and the names
are pointers into the DAG to refer to intermediate expressions. The
{\it leaves} of a shared expression are the leaves of the DAG representing
the shared expression. For the above example, the set of predicates
in $P$ form the leaves of $\tempexpr{a=d}$.

In the rest of the paper, we will illustrate that one can compute a shared 
expression for $\eff_P(e)$ in polynomial or pseudo polynomial time for 
theories that are important for program verification and in particular 
in SLAM. This shared representation can be used to construct 
$\eff_P(e)$ efficiently in practice. 
} 

\comment{
{\bf Remark:} {\it Interestingly on this example, there is only 1 irredundant
derivation for the goal, but our method will not be able to exploit
it. Replace this example with the diamond example? For the diamond example,
with have n preds for n/4 diamonds. Hence there are $2^{n/4}$ derivations. 
But the number of minterms = $2^{n/4}*2^{n/2} = 2^{3n/4}$, as each derivation only
mentions n/2 of the predicates and remaining n/2 can take both positive
and negative values. Thus we also have a compact representation.}

{\bf Remark:} {\it Should we talk about irredundant derivations in this 
paper, because we do not have any way to exploit it. Even our final expressions
are going to contain redundant derivations for this example.}
}
 

\comment{
\subsection{Organization of the paper}
The paper is organized as follows:

\begin{enumerate}
\item Define Symbolic Decision Procedures. 
\item Constructing Symbolic decision procedure for the logic of equality with uninterpreted 
functions and difference logic.
\item Constructing symbolic decision procedures for combination of simple theories 
modularly by using Nelson-Oppen combination result.
\item Handling non-atomic goals. 
\item Implementation and Optimizations.
\item Results on SLAM benchmarks.
\item Conclusions and future challenges.
\end{enumerate}

}

%% file: fastf.tex
\begin{figure}
\begin{boxedminipage}[t]{5.99 truein}
\begin{tabular}{cc}
\begin{minipage}[l]{2.5in}
\infrule
  { X=Y }
  { Y=X }

\vspace{0.1in}

\infrule
  { X=Y \andalso Y=Z }
  { X=Z }
\end{minipage}
& 
\begin{minipage}[l]{2.5in}
\infrule
  { X=Y \andalso X \not= Y }
  { \bot }

\vspace{0.1in}

\infrule
  { X_1=Y_1 \andalso \cdots \andalso X_n=Y_n }
  { f(X_1, \cdots, X_n)=f(Y_1, \cdots, Y_n) }
\end{minipage}
\end{tabular}
\caption{\label{fig:euf_rules} Inference rules for theory of equality and uninterpreted functions.}
\end{boxedminipage}
\end{figure}

\section{Symbolic Decision Procedures (SDP)}
\label{sec:fastf}
We now show how to perform predicate abstraction using symbolic
decision procedures.  We start by describing a saturation-based
decision procedure for a theory $T$ and then use it to describe the
meaning of a symbolic decision procedure for the theory $T$. Finally,
we show how a symbolic decision procedure can yield a shared
expression of $\eff_P(e)$ for predicate abstraction.

A set of predicates $G$ (over theory $T$) is unsatisfiable 
if the formula $(\bigwedge_{g \in G} g)$ is unsatisfiable.
For a given theory $T$, the decision procedure for $T$ 
takes a set of predicates $G$ in the theory and 
checks if $G$ is unsatisfiable. 
A theory is defined by a set of {\it inference rules}.
An inference rule $R$ is of the form:
\infrule[R]
  { A_1 \andalso A_2 \andalso \ldots \andalso A_n }
  { A }
which denotes
that  the predicate $A$ can be derived from predicates $A_1,\ldots,A_n$
in one step.
Each theory has at least one
inference rule for deriving {\it contradiction} ($\bot$).
%
%
We also use $g \derivgoal g_1,\ldots,g_k$ to denote that the predicate
$g$ (or $\bot$, where $g=\bot$) can be derived from the predicates
$g_1,\ldots,g_k$ using one of the inference rules in a single step.
Figure~\ref{fig:euf_rules} describes the inference rules for the
theory of Equality and Uninterpreted Functions.


\subsection{Saturation based decision procedures}
Consider a simple saturation-based procedure $\DP_T$ shown in
Figure~\ref{fig:saturate}, that takes a set of predicates $G$
as input and returns {\sc satisfiable} or {\sc unsatisfiable}.

The algorithm maintains two sets: (i) $W$ is the set of predicates
derived from $G$ up to (and including) the current iteration of the
loop in step (2); (ii) $W'$ is the set of all predicates derived
before the current iteration. These sets are initialized in step (1).
During each iteration of step (2), if a new predicate $g$ can be
derived from a set of predicates $\{g_1,\ldots,g_k\} \subseteq W'$,
then $g$ is added to $W$. The loop terminates after a bound
$\derivDepth{T}{G}$.
%
In step (3), we check if 
{\it any} subset of facts in $W$ can derive contradiction.
If such a subset exists, the algorithm returns 
{\sc unsatisfiable}, otherwise it returns {\sc satisfiable}.

The parameter $d \doteq \derivDepth{T}{G}$ is a bound (that is
determined solely by the set $G$ for the theory $T$) such that if the
loop in step (2) is repeated for at least $d$ steps, then $\DP_T(G)$
returns {\sc unsatisfiable} if and only if $G$ is unsatisfiable. If
such a bound exists for any set of predicates $G$ in the theory, then
$\DP_T$ procedure implements a decision procedure for $T$.

\begin{defi}
  A theory $T$ is called a {\it bounded saturation} theory, if the
  procedure $\DP_T$ described in Figure~\ref{fig:saturate} implements
  a decision procedure for $T$.
\end{defi}

In the rest of the paper, we only consider bounded saturation
theories.  Since there is no ambiguity, we will drop the term
``bounded'' in the rest of the paper and refer to such a theory as
saturation theory.  To show that a theory $T$ is a saturation theory,
it suffices to consider a decision procedure algorithm for $T$ (say
$A_T$) and show that $DP_T$ implements $A_T$. This can be shown by
deriving a bound on $\derivDepth{T}{G}$ for any set $G$ in the theory.

\begin{figure}[t]
\begin{boxedminipage}[t]{5.99 truein}
\begin{enumerate}
\item Initialize $W \leftarrow G$. $W' \leftarrow \{\}$.
\item For $i = 1$ to $\derivDepth{T}{G}$:
\begin{enumerate}
\item Let $W' \leftarrow W$.
\item For every fact $g \not\in W'$, if $(g \derivgoal
  g_1,\ldots,g_k)$ and $g_m \in W'$ for all $m \in [1,k]$:
\begin{itemize}
\item   $W \leftarrow W \cup \{g\}$.
\end{itemize}
\end{enumerate}
\item If ($\bot \derivgoal g_1,\ldots,g_k$) and $g_m \in W$ for all $m \in [1,k]$:
\begin{itemize}
\item return {{\sc unsatisfiable}}
\end{itemize}
\item else return {{\sc satisfiable}}
\end{enumerate}
\indent\caption{\label{fig:saturate} $\DP_T(G)$: A simple saturation-based procedure for theory $T$.
  We use $m \in [i,j]$ to denote $i \leq m \leq j$.}
\end{boxedminipage}\end{figure}



\subsection{Symbolic Decision Procedure}



For a (saturation) theory $T$, a symbolic decision procedure for $T$
($\SDP_T$) takes sets of predicates $G$ and $E$ as inputs, and
symbolically simulates $\DP_T$ on $G' \cup \widetilde{E}$, for every
subset $G' \subseteq G$.  The output of $\SDP_T(G,E)$ is a symbolic
expression representing those subsets $G' \subseteq G$, such that $G'
\cup \widetilde{E}$ is unsatisfiable.  Thus with $|G| = n$, a single
run of $\SDP_T$ symbolically executes $2^n$ runs of $\DP_T$.

We introduce a set of Boolean variables $B_G \doteq \{b_g \; | \; g \in G\}$, 
one for each predicate in $G$. An assignment $\sigma : B_G \rightarrow \{\true,\false\}$ 
over $B_G$ uniquely represents a subset 
$G' \doteq \{g \;| \; \sigma(b_g) = \true \}$ of $G$.

\begin{figure}[t]
\begin{boxedminipage}[t]{5.99 truein}
\begin{enumerate}
\item Initialization 
\begin{enumerate}
\item $W \leftarrow G \cup \widetilde{E}$ and $W' \leftarrow \{\}$.
\item For each $g \in G$, $\tempexpr{(g,0)} \leftarrow b_g$.
\item For each $e_i \in E$, $\tempexpr{(\neg e_i,0)} \leftarrow \true$. 
\end{enumerate}
\item For $i = 1$ to $\maxDerivDepth{T}{G \cup \widetilde{E}}$ do: // Saturation
\begin{enumerate}
\item $W' \leftarrow W$.
\item Initialize $S(g) = \{\}$, for any predicate $g$.
\item For every $g \in W'$, $S(g) \leftarrow S(g) \cup \{\tempexpr{(g,i-1)}\}$.
\item For every $g$, if $(g \derivgoal g_1,\ldots,g_k)$ and $g_m \in W'$ for all $m \in [1,k]$:
\begin{enumerate}
\item Update the set of derivations of $g$ at this level:
\begin{equation}
S(g) \leftarrow S(g) \cup
\{\left( \bigwedge_{m \in [1,k]} \tempexpr{(g_m,i-1)} \right)\}
\label{eqn:temp-expr-update}
\end{equation}
\item $W \leftarrow W \cup \{g\}$.
\end{enumerate}
\item For each $g \in W$:
$\tempexpr{(g,i)} \leftarrow \bigvee_{d \in S(g)} d$
\item For each $g \in W$, $\tempexpr{(g,\top)} \leftarrow \tempexpr{(g,i)}$ 
\end{enumerate}

\item Check for contradiction:
\begin{enumerate}
\item Initialize $S(e) = \{\}$.
\item For every $\{g_1,\ldots,g_k\} \subseteq W$, if $(\bot \derivgoal g_1,\ldots,g_k)$ then
\begin{equation}
S(e) \leftarrow S(e) \cup 
\{\left( \bigwedge_{m \in [1,k]} \tempexpr{(g_m,\top)} \right)\}
\label{eqn:goal-expr-update}
\end{equation}
\item Create the derivations for the goal $e$ as
$\tempexpr{e} \leftarrow  \left(\bigvee_{d \in S(e)} d\right)$
\end{enumerate}
\item Return the shared expression for $\tempexpr{e}$.
\end{enumerate}
\caption{\label{fig:sdp} Symbolic decision procedure $\SDP_T(G,E)$ for theory $T$.
The expression $e$ stands for $\bigvee_{e_i \in E} e_i$.}
\end{boxedminipage}\end{figure}


Figure~\ref{fig:sdp} presents the symbolic decision procedure for a
theory $T$, which symbolically executes the saturation based decision
procedure $\DP_T$ on all possible subsets of the input component $G$.
Just like the $\DP_T$ algorithm, this procedure 
also has three main components: {\it initialization},
{\it saturation} and {\it contradiction} detection.  
The algorithm also maintains sets $W$ and  $W'$, as the $\DP_T$ algorithm does.

Since $\SDP(G,E)$ has to execute $\DP_T(G' \cup \widetilde{E})$ on all
$G' \subseteq G$, the number of steps to iterate the saturation loop equals
the maximum $\derivDepth{T}{G' \cup \widetilde{E}}$ for any
$G' \subseteq G$. For a set of predicates $S$, 
we define the bound $\maxDerivDepth{T}{S}$ as follows:
\[
\maxDerivDepth{T}{S} \doteq  {\it max} \{\derivDepth{T}{S'} \; | \; S' \subseteq S \}
\]

During the execution, the algorithm  constructs a set of shared expressions
with the variables over $B_G$ as the leaves and temporary variables
$\tempexpr{\cdot}$ to name intermediate expressions.  We use
$\tempexpr{(g,i)}$ to denote the expression for the predicate $g$
{\it after} the iteration $i$ of the loop in step (2) of the 
algorithm. We use $\tempexpr{(g,\top)}$ to denote the top-most 
expression for $g$ in the shared expression.
Below, we briefly describe each of the phases of $\SDP_T$:
\begin{enumerate}[:]
\item {\it Initialization} [Step (1)]. 
The set $W$ is initialized to $G \cup \widetilde{E}$ and $W'$ to 
$\{\}$. The leaves of the shared expression symbolically encode each
subset $G' \cup \widetilde{E}$, for every $G' \subseteq G$. 
For each $g \in G$, the leaf $\tempexpr{(g,0)}$ is set to $b_g$. 
For any $e_i \in E$, since $\neg e_i$ is
present in all possible subset $G' \cup \widetilde{E}$,
we replace the leaf for $\neg e_i$ with \true{}.
\item {\it Saturation} [Step (2)]. For each predicate $g$, 
$S(g)$ is the set of derivations of $g$ from predicates in $W'$
during any iteration. For any predicate $g$, we first add all the ways 
to derive $g$ until the previous steps by adding $\tempexpr{(g,i-1)}$ to 
$S(g)$.
Every time $g$ can be derived from some set of facts $g_1,\ldots,g_k$
such that each $g_j$ is in $W'$,  we add this derivation to $S(g)$ in
Equation~\ref{eqn:temp-expr-update}. 
At the end of the iteration $i$, $\tempexpr{(g,i)}$ and $\tempexpr{(g,\top)}$
are updated with the set of derivations in $S(g)$. 
The loop is executed $\maxDerivDepth{T}{G \cup \widetilde{E}}$ times.

\item {\it Contradiction} [Steps (3,4)]. We know that if 
$G' \cup \widetilde{E}$ is unsatisfiable, then $G'$ implies
$e$ (recall, $e$ stands for $\bigvee_{e_i \in E} e_i$). Therefore,
each derivation of $\bot$ from predicates in $W$ gives a 
new derivation of $e$. The set $S(e)$ collects these derivations and
constructs the final expression $\tempexpr{e}$, which
is returned in step (4). 
\end{enumerate}

The output of the procedure is the shared expression $\tempexpr{e}$,
where the leaves of the expression are the variables in $B_G$. The
only operations in $\tempexpr{e}$ are conjunction and disjunction;
$\tempexpr{e}$ is thus a Boolean expression (or a Boolean circuit)
over $B_G$. The internal nodes in the expression are shared and can be
inputs to multiple nodes in the subsequent level.
%
%
%
We now define the evaluation of a (shared) Boolean expression
inductively with respect to a subset $G' \subseteq G$.

\begin{defi}
  For any Boolean expression $\tempexpr{x}$ whose leaves are in set
  $B_G$, and a set $G' \subseteq G$, we define
  $\evalExpr{\tempexpr{x}}{G'}$ as the recursive evaluation of
  $\tempexpr{x}$, after replacing each leaf $b_g$ of $\tempexpr{x}$
  with \true{} if $g \in G'$ and with \false{} otherwise. The
  propositional connectives in the expression ($\wedge$ and $\vee$)
  are interpreted using their standard meaning. 
\label{defn:eval-expr}
\end{defi}

The following theorem explains the correctness of the symbolic
decision procedure.
\begin{thm}
If $\tempexpr{e} \doteq \SDP_T(G,E)$, then for any set of predicates
$G' \subseteq G$, $\evalExpr{\tempexpr{e}}{G'} = \true$ if and only
if $\DP_T(G' \cup \widetilde{E})$ returns {\sc unsatisfiable}.
\label{theorem:symb-dec-proc-correct}
\end{thm}

To prove Theorem~\ref{theorem:symb-dec-proc-correct}, we first
describe an intermediate lemma about $\SDP_T$. To disambiguate between
the data structures used in $\DP_T$ and $\SDP_T$, we use $W_S$ and
$W_S'$ (corresponding to symbolic) to denote $W$ and $W'$ respectively
for the $\SDP$ algorithm. Moreover, it is also clear that $W'$
(respectively $W_S'$) at the iteration $i$ ($i > 1$) is the same as
$W$ (respectively $W_S$) after $i-1$ iterations.


\begin{lem}
For any set of predicates $G' \subseteq G$, at the end of $i$ 
($i\geq 0$) iterations of the loop in step (2) of $\SDP_T(G ,E)$ 
and $\DP_T(G' \cup \widetilde{E})$ procedures:
\begin{enumerate}[\em(1)]
\item $W \subseteq W_S$, and 
\item $\evalExpr{\tempexpr{(g,i)}}{G'} = \true$ if and only if 
$g \in W$ for the $\DP_T$ algorithm.
\end{enumerate}
\label{lemma:sdp-correct-1}
\end{lem}

\begin{proof}
We use an induction on $i$ to prove this lemma, starting from $i = 0$. 

For the base case (after step (1) of both algorithms), 
$W = G' \cup \widetilde{E} \subseteq {G \cup \widetilde{E}} \subseteq W_S$. 
Moreover, for this step,  $\evalExpr{\tempexpr{(g,0)}}{G'}$ for a predicate $g$
can be $\true$ in two ways. 
\begin{enumerate}[(1)]
\item If $g \in \widetilde{E}$, then step (1) of $\SDP_T$
assigns it to \true. Therefore $\evalExpr{\tempexpr{(g,0)}}{G'}$ is \true{} for 
any $G'$. But in step (1) of $\DP_T(G' \cup \widetilde{E})$,
$W$ contains all the predicates in $G' \cup \widetilde{E}$, and therefore
$g \in W$. 
\item If $g \in G'$, then $\evalExpr{\tempexpr{(g,0)}}{G'} = 
\evalExpr{b_g}{G'}$ which is \true, by the definition of $\evalExpr{}{}$. 
Again $g \in W$ after step (1) of the $\DP_T$ algorithm too. 
\end{enumerate}

Let us assume that the inductive hypothesis holds for all values of 
$i$ less than $m$. Consider the iteration number $m$. It is easy to see
that if any fact $g$ is added to $W$ in this step, then $g$ is also added
to $W_S$; therefore part (1) of the lemma is easily established. 

To prove part (2) of the lemma, we will consider
two cases depending of whether a predicate $g$ was present in $W$ before
the $m^{\it th}$ iteration:
\begin{enumerate}[(1)]
\item Let us assume that after $m-1$ iterations of 
$\DP_T(G' \cup \widetilde{E})$ procedure, 
$g \in W$. Since $g$ is never removed from $W$ during any step of
$\DP_T$, $g \in W$ after $m$ iterations too. 
Now, by the inductive hypothesis, 
$\evalExpr{\tempexpr{(g,m-1)}}{G'} = \true$. However, 
$\tempexpr{(g,m-1)} \implies \tempexpr{(g,m)}$ (because 
$\tempexpr{(g,m)}$ contains $\tempexpr{(g,m-1)}$ as one of its
disjuncts in step 2(c) of the $\SDP_T$ algorithm). Therefore,
$\evalExpr{\tempexpr{(g,m)}}{G'} = \true$. 
\item We have to consider two cases depending on whether $g$ can be
derived in $\DP_T(G' \cup \widetilde{E})$ in step $m$. 
\begin{enumerate}[(a)]
\item If $g$ can't be derived in this step in $\DP_T$ algorithm, then
  there is no set $\{g_1,\ldots,g_k\} \subseteq W'$ (of $\DP_T$) such
  that $g \derivgoal g_1,\ldots,g_k$. Since $W'$ is the same as $W$
  after $m-1$ iterations, we can invoke the induction hypothesis to
  show that there exists a predicate $g_j \in \{g_1,\ldots,g_k\}$,
  $\evalExpr{\tempexpr{(g_j,m-1)}}{G'} = \false$. Again, by the
  induction hypothesis, $\evalExpr{\tempexpr{(g,m-1)}}{G'} =
  \false$, since $g \not\in W$ after $m-1$ steps. Thus
  $\evalExpr{\tempexpr{(g,m)}}{G'} = \false$.

\item If $g$ can be derived from $\{g_1,\ldots,g_k\} \subseteq W'$ (of $\DP_T$),
then $\bigwedge_j \tempexpr{(g_j,m-1)}$ implies 
$\tempexpr{(g,m)}$. But for each $g_j \in \{g_1,\ldots,g_k\}$, 
$\evalExpr{(g_j,m-1)}{G'} = \true$ and thus 
$\evalExpr{(g,m)}{G'} = \true$. 
\end{enumerate}
\end{enumerate}
This completes the induction proof. 
\end{proof}

We are now ready to complete the proof of Theorem~\ref{theorem:symb-dec-proc-correct}.
\begin{proof}
Consider the situation where both $\SDP_T(G,E)$ and $\DP_T(G' \cup \widetilde{E})$ have
executed the loop in step (2) for $i = \maxDerivDepth{T}{G \cup \widetilde{E}}$. 
We will consider two cases depending on whether $\bot$ can be derived 
in $\DP_T(G' \cup \widetilde{E})$ in step (3). 
\begin{enumerate}[$\bullet$]
\item Suppose after $i$ iterations, there is a set $\{g_1,\ldots,g_k\} \subseteq W$,
such that $\bot \derivgoal g_1,\ldots,g_k$. This implies that $G' \cup \widetilde{E}$
is unsatisfiable. By Lemma~\ref{lemma:sdp-correct-1}, we know that 
$\evalExpr{\tempexpr{(g_j,\top)}}{G'} = \true$ for each $g_j \in \{g_1,\ldots,g_k\}$, 
and therefore $\evalExpr{\tempexpr{e}}{G'} = \true$. 
\item On the other hand, let $\evalExpr{\tempexpr{e}}{G'} = \true$. This implies 
that there exists a set $\{g_1,\ldots,g_k\} \subseteq W_S$, such that 
$\bot \derivgoal g_1,\ldots,g_k$ and
$\evalExpr{\tempexpr{(g_j,\top)}}{G'} = \true$ for each $g_j \in \{g_1,\ldots,g_k\}$.
By Lemma~\ref{lemma:sdp-correct-1}, we know that $\{g_1,\ldots,g_k\}
\in W$, for the $\DP_T$ procedure too. This means that $DP_T(G' \cup \widetilde{E})$
will return {\sc unsatisfiable}.
\end{enumerate} 
This completes the proof.
\end{proof}

\begin{cor}
For a set of predicates $P$,
if $\tempexpr{e} \doteq \SDP_T(P \cup \widetilde{P},E)$, 
then for any $P' \subseteq (P \cup \widetilde{P})$
representing a minterm over $P$ (i.e. $p_i \in P'$ iff 
$\neg p_i \not\in P'$), 
$\evalExpr{\tempexpr{e}}{P'} = \evalExpr{\eff_P(e)}{P'}$.  
\end{cor}

Hence $\tempexpr{e}$ is a shared expression for $\eff_P(e)$,
where $e$ denotes $\bigvee_{e_i \in E} e_i$.
An explicit representation of 
$\eff_P(e)$ can be obtained by first computing 
$\tempexpr{e} \doteq \SDP_T(P \cup \widetilde{P},E)$
and then enumerating the cubes over $P$ that make $\tempexpr{e}$ \true{}.

In the following sections, we will instantiate $T$ to be the
EUF and DIF theories and show that $\SDP_T$ exists
for such theories. For each theory, we only need to determine the 
value of $\maxDerivDepth{T}{G}$ for any set of predicates $G$.

\begin{exa}
Figure~\ref{fig:sdp_example} demonstrates the working of the
$\SDP(G,E)$ for a simple example. The predicates in $G \doteq \{a=b,
b=c, a=d, d=c\}$ and $E \doteq \{a=c\}$ are limited to equality and
disequality predicates. For this theory $T$, $\maxDerivDepth{T}{G \cup
  \widetilde{E}}$ equals the $\lg(m)$, where $m$ is the number of
terms in $G \cup \widetilde{E}$. We do not show this result for
equality theory in this paper, but prove it for the more general
theory of difference logic in Section~\ref{sec:diff-logic}.
Therefore, we need to iterate Step (2) of the algorithm, for
$lg(\{a,b,c,d\})$ = $2$ steps in Figure~\ref{fig:sdp}.

\begin{figure}[t]
\begin{boxedminipage}[t]{5.99 truein}
\begin{center}
\vskip5 mm
\psfig{figure=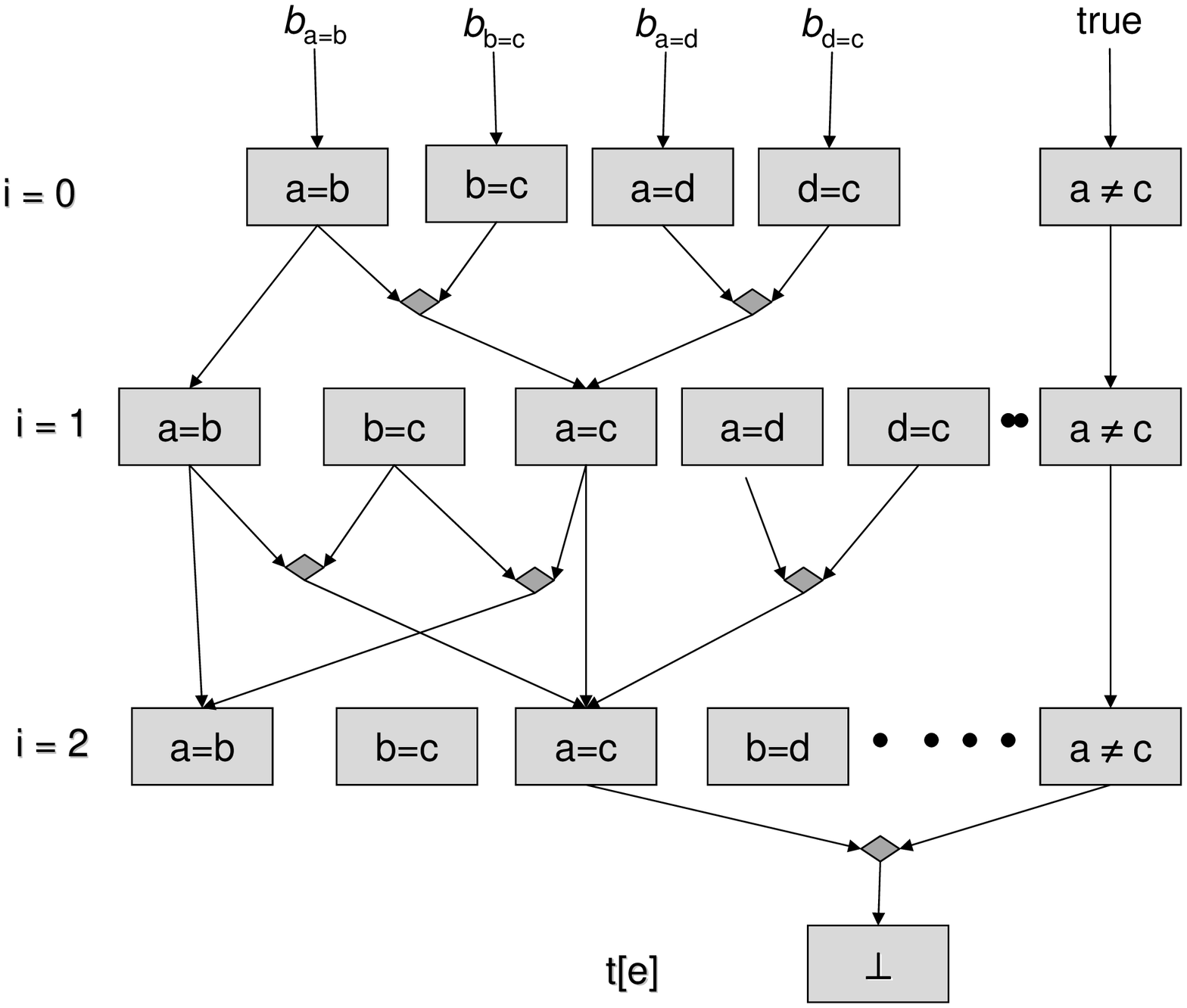,width=3in}
\end{center}
\caption{\label{fig:sdp_example}Example of SDP, where $G \doteq \{a=b,
  b=c, a=d, d=c\}$ and $E \doteq \{a=c\}$. The diamond connective
  represents conjunction, and multiple incoming edges to a node
  represents a disjunction. The node corresponding to the predicate
  $g$ at level $i$ represents $\tempexpr{(g,i)}$. The figure omits
  several nodes and edges at each level to make the diagram readable.}
\end{boxedminipage}\end{figure}

First, a Boolean variable $b_g$ is introduced for each of the
predicate $g \in G$. These variables represent $\tempexpr{(g,0)}$ for
each $g \in G$.  For each $e_i \in \widetilde{E}$, we use $\true$ to
represent $\tempexpr{(e_i,0)}$. Then the Step (2) of the algorithm is
repeated for 2 steps. At each step, new derivations are produced from
the existing set of predicates at the level. The nodes at each level
denotes the set $W$ for the particular iteration. Each derivation
from two predicates in $W$ is represented as the conjunction of the
two predicates (using the diamond connective), and multiple
derivations for a predicate (e.g. 3 ways to derive $a=c$ for $i = 2$)
are represented with multiple incoming edges to a node.

Finally, the contradiction inference rule is used to derive
contradictions ($\bot$) at the last level. Since the only way to
derive contradiction in this example is using $a = c$ and $a \neq c$,
this is the only derivation of $\bot$. The expression $\tempexpr{e}$
represents the acyclic graph rooted at $\bot$, whose leaves are
symbols in $B_G$. The expression $\tempexpr{e}$ intuitively represents
all the derivations of $a=c$ from $G$. More precisely, it represents
all the subsets of $G$ that are inconsistent with $a \neq c$.
\end{exa}

There are a couple of observations that one can make from the previous
example:
\begin{enumerate}[(1)]
\item The expression $\tempexpr{e}$ is a Boolean formula with $B_G$ as
  inputs and an alternation of AND and OR operations. There are no
  negations (NOT) in the formula.
\item Even for this simple example, there are several redundant
  derivations. For example, consider the node $a=b$ in level $i = 2$.
  At this level, $a=b$ can either be derived from $a=b$ or from $b=c$
  and $a=c$, in the previous level. However, the derivation of $a=c$
  in level $i = 1$ already uses $a=b$ (at level $i=0$) for one of its
  derivations. This means that the set of derivations of $a=b$ in
  level $i=2$ contains redundant derivations. These derivations do not
  affect the correctness of the procedure, but simply increases the
  size of $\tempexpr{e}$. However, as we will see in the next two
  sections, the size of the graph for $\tempexpr{e}$ is still (pseudo)
  polynomially bounded for interesting theories.
\end{enumerate}

\begin{rem}
It may be tempting to terminate the loop in step (2) of 
$\SDP_T(G,E)$ once the set of predicates in $W$ does not change 
across two iterations.  However, this would lead to an incomplete 
procedure and the following example demonstrates this.

\begin{exa}
Consider an example where
$G$ contains a set of predicates that denotes an ``almost'' fully
connected graph over vertices $x_1,\ldots,x_n$.  $G$ contains an
equality predicate between every pair of variables except the edge
between $x_1$ and $x_n$.  Let $E \doteq \{x_1 = x_n\}$.

After one iteration
of the $\SDP_T$ algorithm on this example, $W$ will contain an
equality between every pair of variables including $x_1$ and $x_n$
since $x_1 = x_n$ can be derived from $x_1 = x_i, x_i = x_n$, for
every $1 < i < n$. Therefore, if the $\SDP_T$ algorithm terminates
once the set of predicates in $W$ stabilizes, the procedure will 
terminate after two steps. 

Now, consider the subset $G' =
\{x_1=x_2,x_2=x_3,\ldots,x_i=x_{i+1},\ldots,x_{n-1}=x_n\}$ of $G$. 
For this subset of $G$,  $\DP_T(G' \cup \widetilde{E})$ requires 
$lg(n) > 1$ (for $n >2$) steps to
derive the fact $x_1 = x_n$. Therefore $\SDP_T(G,E)$ does not simulate
the action of $\DP_T(G' \cup \widetilde{E})$. More formally,
we can show that $\evalExpr{\tempexpr{e}}{G'} = \false$, but
$G' \cup \widetilde{E}$ is unsatisfiable.
\end{exa}
\end{rem}


\subsection{$\SDP$ for Equality and Uninterpreted Functions}
\label{sec:euf}

\comment{
In this section, we describe the symbolic decision procedure for the 
logic of equality with uninterpreted functions. We will start with the
simple theory of equality between variables (EQ) and then add uninterpreted
functions (EUF).

\subsubsection{Equality Logic}
As noted earlier, a predicate in EQ theory
is of the form $x \sim y$, where $x$ and $y$ are variables and
$\sim \in \{=,\not=\}$. For a set $G$ of EQ predicates,
we define $G_=$ and $G_{\not=}$ to denote the 
set of equality and disequality predicates in $G$.
The inference rules for this
theory {\sc symmetry}, {\sc transitivity} and 
{\sc contradiction} are specified in Figure~\ref{fig:euf_rules}. 
We will assume that the rule of {\sc symmetry} does not generate any new
predicates, as each predicate also denotes its symmetric counterpart.

\begin{prop}
For a set of EQ predicates $G$, if $m$ is the number of variables in $G$,
then $\maxDerivDepth{T}{G}$ for the EQ theory is bound by $lg(m)$. 
\label{prop:eq-max-deriv-depth}
\end{prop}

[Moving the example to the appendix.]

\comment{
One way to prevent generating some  redundant derivations is to maintain for each
predicate $x=y$ in $W$, a set of variables $\neccVars{x=y}$ 
that are {\it necessary} for deriving $x=y$. If $x=y \in G$, then
$\neccVars (x=y)$ is initialized to  $\{x,\}$. During any iteration of the
loop in step (2), when $x=y$ is derived for the first time from predicates 
$x=z,z=y$ in step 2(d), we initialize
$\neccVars{g}$ to $(\bigcup_{m \in [1,k]} \neccVars{g_m}))$. For any
future derivation of $g$ from $g_1,\ldots,g_k$ in step 2(d), we update
$\neccVars{g}$ with $\neccVars{g} \cap (\bigcup_{m \in [1,k]} \neccVars{g_m}))$.
To prevent redundant derivations, for each inference step 
$g \derivgoal g_1,\ldots,g_k$, we add the derivation, only if the variables in 
both the variables in $g$ are not present in $(\bigcup_{m \in [1,k]} \neccVars{g_m}))$.
} 
} 


The terms in this logic can either be variables or 
application of an uninterpreted function symbol to a list of 
terms. 
A predicate in this theory is $t_1 \sim t_2$, where $t_i$ is a 
term and $\sim \;\in \;\{=,\not=\}$. 
For a set $G$ of EUF predicates, $G_=$ and $G_{\not=}$ denote the 
set of equality and disequality predicates in $G$, respectively.
Figure~\ref{fig:euf_rules} 
describes the inference rules for this theory. 

Let ${\it terms}(\phi)$ denote the set of syntactically 
distinct terms in an expression 
(a term or a formula) $\phi$. For example, ${\it terms}(f(h(x)))$
is $\{x, h(x), f(h(x))\}$. For a set of predicates $G$, 
${\it terms}(G)$ denotes the union of the set of terms in any 
$g \in G$. 

A decision procedure for EUF can be
obtained by the {\it congruence closure} algorithm~\cite{nelson-jacm80},
described in Figure~\ref{fig:cong-closure}. 

\begin{figure}[t]
\begin{boxedminipage}[t]{5.99 truein}
\begin{enumerate}
\item Partition the set of 
terms in ${\it terms}(G)$ into equivalence classes using the 
$G_=$ predicates. At any point in the algorithm, let 
$\eClass{t}$ denote the equivalence class for any term
$t \in {\it terms}(G)$.
\begin{enumerate}
\item Initially, each term belongs to its own distinct equivalence
class.
\item We define a procedure ${\it merge}(t_1,t_2)$ that takes two
terms as inputs. The procedure first merges 
the equivalence classes of $t_1$ and $t_2$. If there are two
terms $s_1 \doteq f(u_1,\ldots,u_n)$ and $s_2 \doteq f(v_1,\ldots,v_n)$ such that
$\eClass{u_i} = \eClass{v_i}$, for every $1 \leq i \leq n$, then
it recursively calls ${\it merge}(s_1,s_2)$.
\item For each $t_1 = t_2 \in G_=$, call ${\it merge}(t_1,t_2)$. 
\end{enumerate}
\item If there exists a predicate $t_1 \not= t_2$ in 
$G_{\not=}$, such that $\eClass{t_1} = \eClass{t_2}$, then 
return {\sc unsatisfiable}; else {\sc satisfiable}. 
\end{enumerate}
\caption{\label{fig:cong-closure} Simple description of the congruence closure algorithm.}
\end{boxedminipage}\end{figure}

For a set of predicates $G$, let $m = |{\it terms}(G)|$.
We can show that if we iterate the loop in step (2) of $\DP_T(G)$ 
(shown in Figure~\ref{fig:saturate}) for
at least $3m$ steps, then $\DP_T$ can implement the congruence
closure algorithm. More precisely, for two terms $t_1$ and $t_2$
in ${\it terms}(G)$, the predicate $t_1 = t_2$ will be derived
within $3m$ iterations of the loop in step 2 of $\DP_T(G)$
if and only if $\eClass{t_1} = \eClass{t_2}$ after step (1) of the
congruence closure algorithm (see proof below).

\begin{prop}
For a set of EUF predicates $G$, if $m \doteq
|{\it terms}(G)|$, then the value of 
$\maxDerivDepth{T}{G}$ for the theory is bound by $3m$. 
\label{prop:euf_max_depth}
\end{prop}

\begin{proof}
We first determine the $\derivDepth{T}{G}$ for any set of predicates
in this theory. 

Given a set of EUF predicates $G$, and two terms
$t_1$ and $t_2$ in ${\it terms}(G)$, we need to determine
the maximum number of iterations in step (2) of 
$\DP_T(G)$ to derive $t_1 = t_2$ (if $G_=$ implies
$t_1 = t_2$). 

Recall that the congruence closure algorithm(described in
Figure~\ref{fig:cong-closure}) is a decision procedure for the
theory of EUF. At any point in the algorithm, the terms in $G$
are partitioned into a set of equivalence classes. The operation
$\eClass{t_1} = \eClass{t_2}$ is used to determine if $t_1$ and
$t_2$ belong to the same equivalence class. 

One way to maintain an equivalence class $C \doteq \{t_1,\ldots,t_n\}$
is to keep an equality $t_i = t_j$ between every pair of terms in $C$.
At any point in the congruence closure algorithm, the set of 
equivalence classes corresponds to a set of equalities $C_=$ over terms. 
Then $\eClass{u} = \eClass{v}$ can be implemented by checking
if $u = v \in C_=$. Although this is certainly not an efficient
representation of equivalence classes, this representation allows us
to build $\SDP_T$ for this theory. 

Let us implement the $C_=' \doteq {\it merge}(C_=,t_1,t_2)$ operation that takes
in the current set of equivalence classes $C_=$, two terms $t_1$ and
$t_2$ that are merged and returns the set of equalities $C_='$ denoting the 
new set of equivalence classes. This can be implemented  using the step (2)
of the $\DP_T$ algorithm as follows:
\begin{enumerate}[(1)]
\item $C_=' \leftarrow C_= \cup \{t_1 = t_2\}$.
\item For every term $u \in \eClass{t_1}$, (i.e. $u = t_1 \in C_=$),
add the predicate $u = t_2$ to $C_='$ by
the transitive rule $u = t_2 \derivgoal u = t_1, t_1 = t_2$. Similarly, 
for every $v \in \eClass{t_2}$, add the predicate $v = t_1$ to $C_='$ by 
$v = t_1 \derivgoal v = t_2, t_2 = t_1$. All these steps can be performed
in one iteration of step 2.
\item For every $u \in \eClass{t_1}$ and every $v \in \eClass{t_2}$, 
add the edge $u = v$ to $C_='$ by either of the two transitive rules 
$(u = v \derivgoal u = t_2, t_2 = v)$
or $(u = v \derivgoal u = t_1, t_1 = v)$. 
\item Return $C_='$
\end{enumerate}

If there are $m$ distinct terms in $G$, then there can be at most
$m$ merge operations, as each merge reduces the number of equivalence
classes by one and there were $m$ equivalence classes at the start 
of the congruence closure algorithm. Each merge requires 
three iterations of the step (2) of the $\DP_T$ algorithm to 
generate the new equivalence classes. Hence, we will need at most $3m$ iterations of 
step (2) of $\DP_T$ to derive any fact $t_1 = t_2$ that is implied by
$G_=$.
 
Observe that this decision procedure $\DP_T$ for EUF  
does not need to derive a predicate $t_1 = t_2$ from $G$, if 
both $t_1$ and $t_2$ do not belong to ${\it terms}(G)$. 
Otherwise, if one generates $t_1 = t_2$, then the infinite
sequence of predicates $f(t_1) = f(t_2), f(f(t_1)) = f(f(t_2)),
\ldots$ can be generated without ever converging.

Again, since $\maxDerivDepth{T}{G}$ is the maximum $\derivDepth{T}{G'}$ for any
subset $G' \subseteq G$, and any $G'$ can have at most $m$ terms,
$\maxDerivDepth{T}{G}$ is bounded by $3m$. 
We also believe that
a more refined counting argument can reduce it to $2m$, 
because two equivalent classes can be merged simultaneously in the
$\DP_T$ algorithm. 
\end{proof}

\subsubsection{Complexity of $\SDP_T$}
The run time and size of expression generated by $\SDP_T$ depend both on 
$\maxDerivDepth{T}{G}$ for the theory and also on the maximum number of
predicates in $W$ at any point during the algorithm. 
The maximum number of predicates in $W$ can be at most
$m(m-1)/2$, considering equality between every pair of term. The 
disequalities are never used except for generating contradictions. 
It is also easy to verify that the size of $S(g)$ (used in step (2)
of $\SDP_T$) is  polynomial in the size of input.
Hence the run time of $\SDP_T$ for EUF and the size of the shared
expression returned by the procedure is polynomial in the size of the
input.


\subsection{$\SDP$ for Difference Logic}
\label{sec:diff-logic}

Difference logic is a simple yet useful fragment of linear arithmetic, where
predicates are of the form $x \bowtie y + c$, where $x$, $y$ are variables,
$\bowtie \in \{<,\leq\}$ and $c$ is a real 
constant. Any equality 
$x = y + c$ is represented as a conjunction of $x \leq y + c$ and
$y \leq x - c$.
The variables $x$ and $y$ are interpreted over real numbers. 
The function symbol ``+'' and the
predicate symbols $\{<,\leq\}$ are the interpreted symbols of this theory. %
Figure ~\ref{fig:diff_rules} presents
the inference rules for this 
theory\footnote{Constraints like
  $x \bowtie c$ are handled
  by adding a special variable $x_0$ to denote the 
  constant 0, and rewriting the constraint as 
  $x \bowtie x_0 + c$~\cite{strichman-cav02}.}.

Given a set $G$ of difference logic predicates, we can construct a graph 
where the vertices of the graph are the variables in $G$ and there is a
directed edge in the graph  from $x$ to $y$, labeled with $(\bowtie,c)$ 
if $x \bowtie y + c \in G$. We will use a predicate and
an edge interchangeably in this section. 

\begin{defi}
A simple cycle $x_1 \bowtie x_2 + c_1, x_2 \bowtie x_3 + c_2,
\ldots, x_n \bowtie x_1 + c_n$ (where each $x_i$ is distinct)
is ``illegal'' if the sum of the edges is
$d = \Sigma_{i \in [1,n]} c_i$ and either 
(i) all the edges in the cycle are $\leq$ edges and $d < 0$, or
(ii) at least one edge is an $<$ edge and $d \leq 0$.

\label{defn:illegal-cycle}
\end{defi}

\begin{figure}[t]
\begin{boxedminipage}[t]{5.99 truein}
\begin{tabular}{cc}
\begin{minipage}[l]{2in}
%

\infrule[a]
  { X \leq Z+C \andalso Z \bowtie Y+D}
  { X \bowtie Y+(C+D) }

\vspace{0.1in}

\infrule[b]
  { X < Z+C \andalso Z \bowtie Y+D}
  { X < Y+(C+D) }

\end{minipage}
& 
\begin{minipage}[l]{2.8in}
\infrule[c]
  { X < Y+C \andalso Y \bowtie X+D \andalso C+D \leq 0}
  { \bot }

\vspace{0.1in}

\infrule[d]
  { X \leq Y+C \andalso Y \leq X+D \andalso C+D < 0}
  { \bot }

\vspace{0.1in}

\infrule[e]
  { X \leq Y \andalso Y \leq X}
  { X = Y }
\end{minipage}
\end{tabular}
\caption{\label{fig:diff_rules} Inference rules for Difference logic.}
\end{boxedminipage}\end{figure}

It is well known~\cite{cormen-book} that a set of difference predicates
$G$ is unsatisfiable if and only the graph constructed from the 
predicates has a simple illegal cycle. 
Alternately, if we add an edge $(\bowtie,c)$ between $x$ and $y$
for every simple path from $x$ to $y$ of weight $c$ ($\bowtie$ determined
by the labels of the edges in the path),
then we only need to check for simple cycles of length two in the
resultant graph. This corresponds to the rules (C) and (D) in
Figure~\ref{fig:diff_rules}.

For a set of predicates $G$, a predicate corresponding 
to a simple path in the graph of $G$ can be derived within 
$lg(m)$ iterations of step (2) of $\DP_T$ procedure, 
where $m$ is the number of variables in $G$ (see proof below). 

\begin{prop}
For a set of DIF predicates $G$, if $m$ is the number of variables in $G$,
then $\maxDerivDepth{T}{G}$ for the DIF theory is bound by $lg(m)$. 
\label{prop:dif-max-deriv-depth}
\end{prop}

\begin{proof}
\comment{
Consider the algorithm $\DP_T(G)$ for this theory. Let $m$ be the number of
variables in $G$. Let $W$ be the set of facts derived from $G$ after
$lg(m)$ steps of the algorithm $\DP_T(G)$.
We claim that to detect any negative weight cycle in $G$, 
it suffices to check if there are two edges 
$x \bowtie y + c$ and $y \bowtie x + d$ in $W$, such that they form an
illegal cycle. Thus $\maxDerivDepth{T}{G}$ for this theory is 
also $lg(m)$, $m$ being the number of variables in $G$. We briefly
explain the reason here.
}

It is not hard to see that if there is a simple path 
$x \bowtie_1 x_1 + c_1, x_1 \bowtie_2 x_2 + c_2,\ldots, x_{n-1} \bowtie_{n} y + c_n$
in the original graph of $G$, then  after $lg(m)$ iterations of
the loop in step (2), there is a predicate $x \bowtie' y + c$ in $W$; where
$c = \Sigma_{i\in [1,n-1]} c_i$ and $\bowtie'$ is $<$ if at least 
one of $\bowtie_i$  is $<$ and $\leq$ otherwise. 
This is because
if there is a simple path between $x$ and 
$y$ through edges in $G$ with length (number of edges from $G$) 
between $2^{i-1}$ and $2^i$, then
the algorithm  $\DP_T$ generates a predicate for the path during
iteration $i$.

However, $\DP_T$ can produce a predicate
$x \bowtie y + c$, even though none of the simple paths between 
$x$ and $y$ add up to this predicate. 
These facts are generated by the non-simple paths that go around cycles 
one or more times. Consider the 
set $G \doteq \{x<y+1, y<x-2, x < z - 1,\ldots\}$. In this case we can produce
the fact $y < z - 3$ from $y < x - 2, x < z-1 $ and then $x < z -2$
from $y < z -3, x < y + 1$. 

To prove the correctness of the $\DP_T$ algorithm, we 
will show these additional facts can be safely generated. 
Consider two cases:
\begin{enumerate}[$\bullet$]
\item Suppose there is an illegal cycle in the graph. In that case, after
$lg(m)$ steps, we will have two facts $x \bowtie y + c$ and  $y \bowtie x + d$ 
in $W$ such that they form an illegal cycle. Thus $\DP_T$ returns unsatisfiable.

\item Suppose there are no illegal cycles in the original graph for $G$.
For simplicity, let us assume that there are only $<$ edges in the graph. 
A similar argument can be made when $\leq$ edges are present. 

In this case, every cycle in the graph has a strictly positive weight. 
A predicate $x \bowtie y + d$ can be generated from non-simple paths only if
there is a predicate $x \bowtie y + c \in G$ such that $c < d$.
The predicate $x \bowtie y + d$ can't be a part of an illegal cycle, because
otherwise $x \bowtie y + c$ would have to be part of an illegal cycle too.
Hence $\DP_T$ returns satisfiable. 
\end{enumerate}

Note that we do not need any inference rule to weaken a predicate,
$X < Y + D \derivgoal X < Y + C$, with $C < D$.  This is because we use the
predicates generated only to detect illegal cycles. 
If a predicate $x < y + c$ does not form an illegal cycle, then neither does 
any weaker predicate $x < y + d$, where $d \geq c$.
\end{proof}


\subsubsection{Complexity of $\SDP_T$}
Let $c_{\it max}$ be the
absolute value of the largest constant in the set $G$. 
We can ignore any derived predicate in of the form $x \bowtie y + C$ from the set $W$
where the absolute value of $C$ is greater than $(m-1)*c_{\it max}$. 
This is because the maximum weight of any simple path between 
$x$ and $y$ can be at most $(m-1)*c_{\it max}$. 
Again, let {\it const($g$)} be the absolute value of the 
constant in a predicate $g$. The maximum weight on any simple path has to
be a combination of these weights. Thus, the absolute value of the constant
is bound by:
$$
C \leq min\{(m-1)*c_{\it max},\Sigma_{g \in G} {\it const(g)}\}
$$

The maximum number of derived predicates in $W$ can be $2*m^2*(2*C+1)$, where a
predicate can be either $\leq$ or $<$, with $m^2$ possible variable pairs
and the absolute value of the constant is bound by $C$. This is a {\it pseudo
polynomial} bound as it depends on the value of the constants in the input. 

However, many program verification queries use a subset of difference logic
where each predicate is of the form $x \bowtie y$ or $x \bowtie c$. 
For this case, the 
maximum number of predicates generated can be $2*m*(m-1+k)$, where 
$k$ is the number of different constants in the input.  

\section{Combining $\SDP$ for saturation theories}
\label{sec:combine-sdp}
In this section, we  provide a method to construct a symbolic decision
procedure for the combination of saturation theories $T_1$ and $T_2$, given 
$\SDP$ for $T_1$ and $T_2$. The combination is based on an extension of 
the Nelson-Oppen (N-O) framework~\cite{nelson-toplas79} 
that constructs a decision  procedure for the theory $T_1 \cup T_2$ using the 
decision procedures of $T_1$ and $T_2$.

We assume that the theories $T_1$ and $T_2$ have disjoint signatures (i.e.,
they do not share any function symbol), and each theory $T_i$ is 
{\it convex} and {\it stably infinite}\footnote{We need these 
restrictions only to exploit the N-O combination result.
The definition of convexity and stably infiniteness
can be found in~\cite{nelson-toplas79}.}.
Let us briefly explain the N-O method for combining decision
procedures before explaining the method for combining $\SDP$. 

\subsection{Nelson-Oppen method for Combining Decision Procedures} 
\label{sec:nelson-oppen}

Given two theories $T_1$ and $T_2$, and the decision procedures
$\DP_{T_1}$ and $\DP_{T_2}$, the N-O framework 
constructs the decision procedure for $T_1 \cup T_2$,
denoted as $\DP_{T_1 \cup T_2}$. 

To decide an input set $G$, the first step in the procedure is to {\it purify}
$G$ into sets $G_1$ and $G_2$ such that $G_i$ only contains symbols 
from theory $T_i$
and $G$ is satisfiable if and only if $G_1 \cup G_2$ is satisfiable. 
Consider a predicate $g \doteq p(t_1,\ldots,t_n)$ in $G$, where $p$ is
a theory $T_1$ symbol. The predicate $g$ is purified to $g'$ by replacing 
each subterm $t_j$ whose top-level symbol does not 
belong to $T_1$ with a fresh variable $w_j$. 
The expression $t_j$ is then purified to $t_j'$ recursively.
We add $g'$ to $G_1$ and the {\it binding predicate} 
$w_j = t_j'$ to the set $G_2$. We denote the latter as binding
predicate because it binds the fresh variable $w_j$ to a term
$t_j'$.  

Let $\varset{sh}$ be the set of {\it shared} 
variables that appear in $G_1 \cap G_2$.
A set of equalities  $\Delta$ over variables in $\varset{sh}$ is 
maintained; $\Delta$ records the set of equalities implied by 
the facts from either theory. Initially, $\Delta = \{\}$.

Each theory $T_i$ then alternately decides 
if $\DP_{T_i}(G_i \cup \Delta)$  is unsatisfiable.
If any theory reports {\sc unsatisfiable},
the algorithm returns {\sc unsatisfiable}; otherwise, 
the theory $T_i$ generates the new set of equalities over $\varset{sh}$
that are implied by $G_i \cup \Delta$\footnote{We assume that 
each theory has an inference rule for deriving equality between
variables in the theory, and $\DP_T$ also returns a set of equality
over variables.}.  
These equalities are added to $\Delta$ and are 
communicated to the other theory. 
This process is continued until the set $\Delta$ does not change.
In this case, the method returns {\sc satisfiable}. Let us
denote this algorithm as $\DP_{T_1 \cup T_2}$.


\begin{thm}[\cite{nelson-toplas79}]
For convex, stably infinite  and signature-disjoint theories $T_1$ and $T_2$, 
$\DP_{T_1 \cup T_2}$ is a decision procedure for $T_1 \cup T_2$.
\end{thm}

There can be at most $|\varset{sh}|$ irredundant equalities over 
$\varset{sh}$, therefore the N-O loop terminates after 
$|\varset{sh}|$ iterations for any input. 

\comment{ 
We will use this observation to define the $\derivDepth{T_1 \cup T_2}
{G}$ for any set of predicates $G$ for the combined theory. 

\begin{prop}
Let $\Delta_{sh}$ be any subset of equalities over $\varset{sh}$. 
For any set of predicates $G$ over $T_1 \cup T_2$, $\derivDepth{T_1 \cup T_2}{G}$
is bound by the maximum value of $|\varset{sh}|/2 * (\derivDepth{T_1}{G_1 \cup \Delta_{sh}}
+ \derivDepth{T_2}{G_2 \cup \Delta_{sh}})$, for any $\Delta_{sh}$.
\label{prop:max-no-depth}
\end{prop}

For the combined theory of EUF ($T_1$) and DIF ($T_2$), the $\derivDepth{T_1 \cup T_2}{G}$
reduces to $|\varset{sh}|/2 * (m + lg(n))$, where $m$ is the
${\it terms}(G_1)$ and $n$ is number of variables in $G_2$. This is because 
$\Delta_{sh}$ never increases the number of terms or variables in 
either theory, and the $\derivDepth{}{G}$ for each theory only 
depends on the number of terms or variables in $G$.
}



\subsection{Combining $\SDP$ using Nelson-Oppen method}
\label{sec:nelson-oppen-sdp}

We will briefly describe a method to construct the 
$\SDP_{T_1 \cup T_2}$ by combining $\SDP_{T_1}$ and $\SDP_{T_2}$. 
As before, the input to the method is the pair $(G,E)$ and the 
output is an expression $\tempexpr{e}$. The facts in $E$ are 
also purified into sets $E_1$ and $E_2$ and the new binding 
predicates are added to either $G_1$ or $G_2$. 

Our goal is to
symbolically encode  the runs of the N-O procedure for
$G' \cup \widetilde{E}$, for every $G' \subseteq G$. 
For any equality predicate $\delta$ over $\varset{sh}$, we maintain
an expression $\psi_\delta$ that records all the different
ways to derive $\delta$ (initialized to \false{}). 
We also maintain an expression $\psi_e$ to record all the derivations of 
$e$ (initialized to \false{}).

The N-O loop operates just like the case for constructing
$\DP_{T_1 \cup T_2}$. 
The $\SDP_{T_i}$ for each theory $T_i$ now takes 
$(G_i \cup \Delta, E_i)$ as input, where $\Delta$ is 
the set of equalities over $\varset{sh}$ derived so far.
In addition to computing the (shared) expression 
$\tempexpr{e}$ as before, $\SDP_{T_i}$ also returns the expression
$\tempexpr{(\delta,\top)}$, for each equality $\delta$ 
over $\varset{sh}$ that can be derived in
step (2) of the $\SDP_T$ algorithm.

The leaves of the expressions $\tempexpr{e}$ and
$\tempexpr{(\delta,\top)}$ are $G_i \cup \Delta$ (since leaves for
$\widetilde{E_i}$ are replaced with \true{}). We substitute the leaves
for any $\delta \in \Delta$ with the expression $\psi_\delta$, to
incorporate the derivations of $\delta$ until this point.  We also
update $\psi_\delta \leftarrow \left(\psi_\delta \vee
\tempexpr{(\delta,\top)}\right)$ to add the new derivations of
$\delta$.  Similarly, we update $\psi_e \leftarrow \left(\psi_e \vee
\tempexpr{e}\right)$ with the new derivations.

The N-O loop iterates $|\varset{sh}|$ number of times to ensure
that it has seen every derivation of a shared equality over 
$\varset{sh}$ from any set
$G_1' \cup G_2' \cup \widetilde{E_1} \cup \widetilde{E_2}$,
where $G_i' \subseteq G_i$. 

After the N-O iteration terminates, $\psi_e$ contains all the derivations
of $e$ from $G$. However, at this point, there are two kind of 
predicates in the leaves of $\psi_e$; the purified predicates
and the binding predicates. 
If $g'$ was the purified form of a predicate
$g \in G$, we replace the leaf for $g'$ with $b_g$. 
The leaves of the binding predicates are replaced with $\true$,
as the fresh variables in these predicates are really names for subterms
in any predicate, and thus their presence does not affect the satisfiability
of a formula. 
Let $\tempexpr{e}$ denote the final expression for $\psi_e$ that is
returned by $\SDP_{T_1 \cup T_2}$. 
Observe that the leaves of $\tempexpr{e}$ are variables in $B_G$.

\comment{
Before we proceed, let us define a few conventions: 
Since $\SDP_T$ for a theory $T$ is invoked multiple times within the 
N-O loop, 
the internal expression names in $\SDP_T$ look like $\tempexpr{(g,l,i)}$ where $g$ is the
predicate, $l$ is the iteration number of N-O loop and $i$ is the
iteration number of the internal loop in $\SDP_T$. 
Similarly, $\tempexpr{(g,l,\top)}$ refers to the top-level expression for 
$g$, during the iteration $l$ of the N-O loop.  For a set of predicates $H$, 
we also define $\tempexprset{H}$ to be a set of expression names, one for each 
$h \in H$.

Each $\SDP_{T_i}$ takes as input the sets $G_i$, $E_i$,
a set of expression names 
$\tempexprset{\Delta}$ (one  for each shared equality
in $\Delta$) and the iteration number $l$. It returns the expression for 
$\tempexpr{e}$, the new set of shared equalities $\Delta$ and the expression
$\tempexpr{(\delta,l,\top)}$ for each shared equality $\delta \in \Delta$.

\begin{figure}[t]
\begin{boxedminipage}[t]{5.99 truein}
\begin{enumerate}
\item Let $(G_1,G_2,E_1,E_2) \leftarrow {\it Purify}(G,E)$. Let $G^p_i \subseteq
G_i$ be the set of predicates that bind the fresh variables to their
corresponding terms in the two theories. 
\item Initialize $\Delta \leftarrow \{\}$.
Make $\tempexpr{\Delta} \leftarrow \{\}$.  
\item For $i = 1$ to $|\varset{sh}|/2 + 1$ do:
\begin{enumerate}
\item Let $(\tempexpr{e}^1,\Delta,\tempexprset{\Delta})
\leftarrow \SDP_{T_1} (G_1,E_1,\tempexprset{\Delta},2i-1)$;
\item Let $(\tempexpr{e}^2,\Delta,\tempexprset{\Delta})
\leftarrow \SDP_{T_2} (G_2,E_2,\tempexprset{\Delta},2i)$;
\end{enumerate}
\item Assign $\tempexpr{e} \leftarrow \tempexpr{e}^1 \vee \tempexpr{e}^2$.
\item Return $\tempexpr{e}$.
\end{enumerate}
\caption{\label{fig:combine_sdp} $\SDP_T$ for a $T_1 \cup T_2$.}
\end{boxedminipage}\end{figure}

Figure~\ref{fig:combine_sdp} describes the $\SDP_T$ algorithm for the
theory $T_1 \cup T_2$.
Each theory first computes $\SDP_{T_i}(G_i \cup \Delta, E_i)$
(exactly similar to the description in Figure~\ref{fig:sdp})
and then replaces the leaf for each $\delta \in \Delta$, with the
expression for $\delta$ from $\tempexprset{\Delta}$.
[Needs slightly better way to present this.]
It then  communicates the new expressions 
for the set of equalities in $\Delta$ to the other theory. 
This process is repeated for the maximum number of times the
N-O procedure can execute $\DP_{T_1 \cup T_2}$
on any subset of the inputs. For the final iteration, 
we generate the expression for $\tempexpr{e}$ from each theory
and return the disjunction of the two expressions. This is because
contradiction can be derived in both the theories .

The leaves of the expression $\tempexpr{e}$ correspond to predicates
in $G_1 \cup G_2 \cup E_1 \cup E_2$. Since the predicates in 
$E_1 \cup E_2$ are present in every run of $\DP_T(G' \cup \widetilde{E})$,
for any $G' \subseteq G$, we can replace these leaves by $\true$ as 
before. A leaf that corresponds to $g'$ (the purified version of a 
predicate $g \in G$) is replaced with $b_g$. Finally, a leaf that
corresponds to a binding predicate $h \in G^p_i$ 
also is replaced with $\true$. 
Although the set of binding predicates for $G' \cup \widetilde{E}$
depend on the specific subset $G' \subseteq G$ that is purified, 
we can show that adding all the binding predicates that result
from purifying $G \cup \widetilde{E}$ does
not affect the correctness of $\DP_T$ algorithm. Therefore, 
the leaves of $\tempexpr{e}$ are Boolean variables in $B_G$.

}
\begin{thm}
For two convex, stably-infinite and signature-disjoint theories $T_1$ and 
$T_2$, if $\tempexpr{e} \doteq \SDP_{T_1 \cup T_2}(G,E)$, 
then for any set of predicates
$G' \subseteq G$, $\evalExpr{\tempexpr{e}}{G'} = \true$ if and only
if $\DP_{T_1 \cup T_2}(G' \cup \widetilde{E})$ returns {\sc unsatisfiable}.
\label{theorem:sdp-combine}
\end{thm}
Since the theory of EUF and DIF satisfy all the 
restrictions of the theories of this section, we can
construct an $\SDP$ for the combined theory that still
runs in pseudo-polynomial time.

%% file: impl.tex
\section{Implementation and Results}
\label{sec:impl}
We have implemented a prototype of the symbolic decision procedure for
the combination of EUF and DIF theories.
To construct $\eff_P(e)$, we first build a BDD
(using the CUDD~\cite{cudd-www} BDD package) for the expression
$\tempexpr{e}$ (returned by $\SDP_T(P \cup \widetilde{P}, E)$) and
then enumerate the cubes from the BDD.

Creating the BDD for the shared expression $\tempexpr{e}$ 
and enumerating the cubes from the BDD can have exponential
complexity in the worst case. This is because the expression
for $\eff_P(e)$ can involve an exponential number of cubes (e.g.
the example in Fig~\ref{tab:diamond-results}). However, most problems 
in practice have a few cubes in $\eff_P(e)$.
%
Secondly, as the number of
leaves of $\tempexpr{e}$ (alternately, number of BDD variables) 
is bound by $|P|$, the size of the overall BDD is usually small, 
and is computed efficiently in practice. 
Finally, by generating only the {\it prime implicants}\footnote{
For any Boolean formula $\phi$ over variables in $V$, 
prime implicants of $\phi$ is a set of cubes $C \doteq \{c_1,\ldots,c_m\}$
over $V$ such that $\phi \Leftrightarrow \bigvee_{c \in C} c$ and 
two or more cubes from $C$ can't be combined to form a larger cube.} 
of $\eff_P(e)$ from the BDD,
we obtain a compact representation of $\eff_P(e)$.

\comment{

\subsection{Constructing Shared Boolean Programs}
The symbolic decision procedure can be used to generate
a shared representation of $\eff_P(e)$. This shared
expression can be used to construct a Boolean Program
directly, where the Boolean program allows 
{\tt let t\_\{e1\} = e2 in e3} constructs to store intermediate 
expressions. The intermediate variables {\tt t\_\{e1\}} act as
additional state variables in the Boolean Program. This pushes
the complexity of predicate abstraction phase (that has often
proven to be the bottleneck in many predicate abstraction based 
tools including SLAM) to the Boolean model checking
phase. In order to make this approach feasible, we will need 
heuristics to reduce the number of temporary variables generated 
during the process to add fewer variables to the Boolean Program. 
We plan to evaluate this tradeoff on real C programs soon.

\subsection{BDD-based construction of $\eff_P(e)$}
We currently use Binary Decision Diagrams to construct the
expression for $\eff_P(e)$ from the shared expression 
$\tempexpr{e}$ returned by $\SDP_T(P \cup \bar {P},e)$. 
We construct a BDD for the 
expression $\tempexpr{e}$ by recursively computing the BDDs
for the subexpressions of $\tempexpr{e}$. Since the number of
leaves of the expression (bounded by $|G|$) is usually small 
($<50$) for any reasonably sized predicate abstraction query, 
the size of the overall BDD is manageable and the BDD is computed
efficiently in practice. 

There are many advantages of using BDD to construct the final expression
for $\eff_P(e)$. First, during evaluating $\tempexpr{e}$, we need
to store the set of derivations for each intermediate node in the 
expression, to avoid recomputation. 
Storing this set explicitly at each node will incur 
a blowup in space in practice. However, since each node now only 
maintains a pointer to the BDD representation of its derivation, 
we achieve sharing across the expressions for all the nodes. Secondly, 
to construct $\eff_P(e)$, we enumerate the set of 
{\it prime-implicants} from the BDD for $\tempexpr{e}$ ({\bf Explain}).
This generates a compact representation of $\eff_P(e)$ in many cases.
} 


\input{results}


\comment{
\subsection{Approximations}
\label{sec:approx}

As described in Section~\ref{sec:setup}, computing $\eff_P(e)$ for an
arbitrary expressions $e$ require us to construct a CNF equivalent of
$e$. This might result in an exponential blowup on many examples. 
However, we can construct a conservative approximation of $\eff_P(e)$
by distributing $\eff_P(e)$ over disjunctions as well: i.e., 
compute $\eff_P(e_1 \vee e_2)$ by computing $\eff_P(e_1) \vee
\eff_P(e_2)$. By Proposition~\ref{prop:F_properties}, this yields
an underapproximation of $\eff_P(e)$, but it may suffice in 
practice. 

Pushing through disjunctions. Why suffices for Boolean Program
construction. Constrain loop. 
} 

\comment{
\begin{itemize}


\item Incremental nature of the main derivations and use for multiple
parts of an expression or for updating different predicates.

\item Depth of derivation is lg(N) for pure EQ facts and DIFF facts. 
Also when congruence is not triggered within lg(D) steps, we know that 
congruence can't be fired after that. In fact for cases with uninterpreted
functions, we might have only few  between  terms which have a lot
of subterms; in this case if congruence rule is not triggered after 
lg(|E|) depth,  we can stop. 

\item Very few shared terms for SLAM examples. Alternation only a 
small constant number of times. (Statistics). 

\item Distinct facts from SLAM.
\end{itemize}
}

%% file: results.tex
\begin{figure}[t]
\begin{minipage}{2.5in}
\begin{center}
\psfig{figure=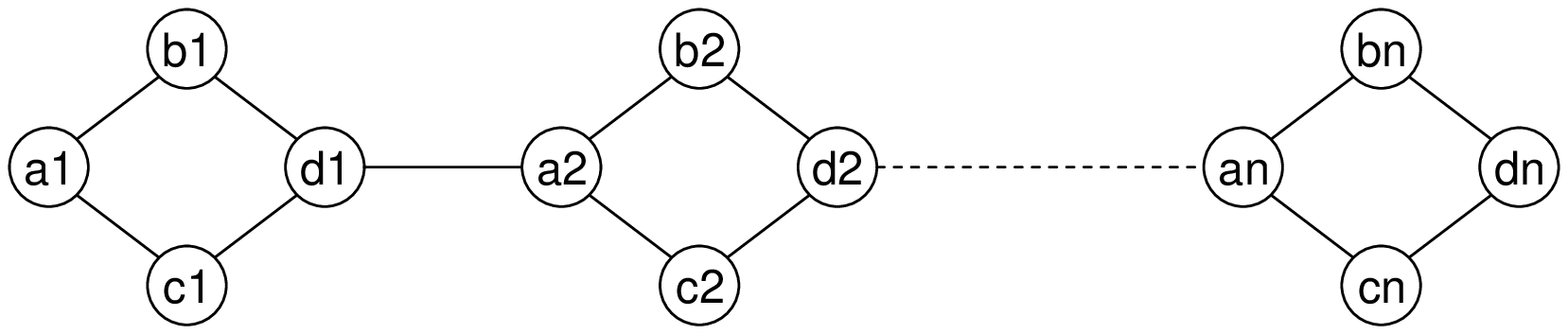,width=.55in,angle=270}
\end{center}
\end{minipage}
\begin{minipage}{2.5in}
\begin{center}
\begin{tabular}{||c|c|c|c||}
\hline
~~$n$~~ &~~ $|P|$~~& ~~$\SDP_T$~~  & ~~{\sc UCLID}~~ \\
& & time (s) & time (s) \\
\hline
3 & 14 & 0.20 & 19.37 \\ \hline
4 & 19 & 0.43 & 656 \\ \hline
5 & 24 & 0.65 & -  \\ \hline
10 & 49 & 5.81 & - \\ \hline
12 & 59 & 12.28 & - \\ \hline
\end{tabular}
\end{center}
\end{minipage}
\caption{\label{tab:diamond-results} Result on diamond examples with increasing number of diamonds.
The expression $e$ is $(a1 = dn)$. A ``-'' denotes a timeout of 1000 seconds.}
\end{figure}


We report preliminary results evaluating our symbolic decision
procedure based predicate abstraction method on a set of software
verification benchmarks.  The benchmarks are generated from the
predicate abstraction step for constructing Boolean Programs from C
programs of Microsoft Windows device drivers in
SLAM~\cite{ball-pldi01}.

We compare our method with two other methods for performing predicate
abstraction:
\begin{description}
\item {{\sc DP}-based}: This method uses the decision procedure {\sc
    zapato}~\cite{ball-cav04} to enumerate the set of cubes that imply
  $e$. Various optimizations (e.g. considering cubes in increasing
  order of size) are used to prevent enumerating exponential number of
  cubes in practice.
\item  {{\sc UCLID}-based}: 
  This method performs quantifier-elimination using incremental
  SAT-based methods~\cite{lahiri-cav03a}. The procedure works by first
  converting the problem into an existential quantifier elimination
  problem in first-order logic and then reducing it to Boolean
  quantifier elimination by using an encoding to Boolean logic.
  Finally, it uses SAT-based methods for performing Boolean
  quantification.
\end{description}

To compare with the {\sc DP}-based method, we generated 665 predicate
abstraction queries from the verification of device-driver programs.
Most of these queries had between 5 and 14 predicates in them and are
fairly representative of queries in SLAM.  The run time of {\sc
  DP}-based method was 27904 seconds on a 3 GHz. machine with 1GB
memory. The run time of $\SDP$-based method was 273 seconds. This
gives a little more than 100X speedup on these examples, demonstrating
that our approach can scale much better than decision procedure based
methods.
We have not been able to run {\sc UCLID}-based method on these
particular SLAM benchmarks; the {\sc UCLID}-based tool is no longer
actively maintained, and we had trouble translating these SLAM
benchmarks to input of UCLID. From our earlier experience of using
{\sc UCLID} on similar benchmarks (Fig. 3 in ~\cite{lahiri-cav03a}),
we believe that most of these benchmarks can be solved within a few
seconds, and the total runtime would not differ by more than 2--3X (in
favor of the current technique).

To compare with {\sc UCLID}-based approach, we generated different
instances of a problem (see Figure~\ref{tab:diamond-results} for the
example) where $P$ is a set of equality predicates representing $n$
diamonds connected in a chain and $e$ is an equality $a1 = dn$. We
generated different problem instances by varying the size of $n$. For
an instance with $n$ diamonds, there are $5n-1$ predicates in $P$ and
$2^n$ cubes in $\eff_P(e)$ to denote all the paths from $a1$ to $dn$.
Figure~\ref{tab:diamond-results} shows the result comparing both the
methods. We should note that {\sc UCLID} method was run on a slightly
slower 2GHz machine.  The results illustrate that our method scales
much better than the SAT-based enumeration used in UCLID for this
example.  Intuitively, UCLID-based approach grows exponentially with
the number of predicates ($2^{|P|}$), whereas our approach only grows
exponentially with the number of diamonds ($2^n$) in the result.



%% file: concl.tex
\section{Conclusions and future work}
\label{sec:concl}

In this paper, we have presented the concept of symbolic decision
procedures and showed its use for predicate abstraction. We have
provided an algorithm for synthesizing a SDP for any bounded
saturation theory. We show that such SDP exists for interesting
theories such as EUF and difference logic. These SDP construct a
shared expression and run with polynomial and pseudo-polynomial
complexity respectively. Finally, we have provided a method for
constructing the SDP for simple mixed theories using an extension of
the Nelson-Oppen combination framework. Preliminary results comparing
it some of the existing approaches are encouraging.

There are several avenues of future work, some of which are outlined
 below:
\begin{enumerate}[$\bullet$]
\item First, it is interesting to find out how to construct a SDP for
  other theories, including the theory of linear arithmetic (over
  rationals). For linear arithmetic, one can perform a ``symbolic''
  Fourier-Motzkin~\cite{fourier-motzkin} elimination procedure to
  construct an SDP --- the inference rule would eliminate a variable
  from all the predicates in a given level.  However, it is not clear
  how to generate implied equalities from such a procedure to combine
  the SDP with SDP for other theories.
  
\item Second, as the example in Figure~\ref{fig:sdp_example}
  illustrated, there are a lot of redundant derivations present in the
  resultant expression. The algorithm will benefit from optimizations
  that can minimize such redundant derivations.

\item Extend the combination of SDPs to non-convex theories.   

\end{enumerate}